\documentclass[a4paper]{article}
\usepackage{amsfonts,amssymb,amsmath,amsthm,cite}
\usepackage{ucs} 
\usepackage[utf8x]{inputenc}
\usepackage{graphicx}
\usepackage[english]{babel}
\usepackage{slashed}
\usepackage{textcomp}
\usepackage{bm}
\usepackage{titlesec}
\usepackage{mathtools}
\usepackage{etoolbox}
\patchcmd{\thebibliography}{\section*}{\section}{}{}

\evensidemargin=\oddsidemargin
\newtheorem{theorem}{Theorem}

\begin{document}
\vspace{10mm}
\begin{center}
	\large{\textbf{Three-loop renormalization of the quantum action}\\ 
		\textbf{for a five-dimensional scalar cubic model with the usage}\\
		\textbf{of the background field method and a cutoff regularization}}
\end{center}
\vspace{2mm}
\begin{center}
	\large{\textbf{A. V. Ivanov~~~~~N. V. Kharuk}}
\end{center}
\begin{center}
St. Petersburg Department of Steklov Mathematical Institute of Russian Academy of Sciences,\\ 
27 Fontanka, St. Petersburg 191023, Russia
\end{center}
\begin{center}
Leonhard Euler International Mathematical Institute in Saint Petersburg,\\ 
10 Pesochnaya nab., St. Petersburg 197022, Russia
\end{center}
\begin{center}
	E-mail: regul1@mail.ru
\end{center}
\begin{center}
	E-mail: natakharuk@mail.ru
\end{center}
\vspace{2mm}
\begin{flushright}
	\large{\textbf{\textit{To the 90-th anniversary of L.D.Faddeev}}}
\end{flushright}
\vspace{10mm}

\textbf{Abstract.} The paper studies the quantum action for the five-dimensional real $\phi^3$-theory in the case of a general formulation using the background field method. The three-loop renormalization is performed with the usage of a cutoff regularization in the coordinate representation. The explicit form of the first three coefficients for the renormalization constants is presented. The absence of non-local singular contributions and partial results for the fourth correction are discussed.

\vspace{2mm}

\newpage	

\tableofcontents
\section{Introduction}
\label{30:sec:int}

Regularization occupies an important position in quantum field theory \cite{9,10}, especially when it comes to the perturbative approach, which often produces divergent integrals \cite{6,7}. Unfortunately, there is no universal recipe in the regularization theory that would be equally good for all models. This is primarily due to the fact that a regularization is directly related to the deformation of the classical action, which necessarily leads to the loss of certain properties of the studied model. This position makes the theory of regularization an actual and in-demand area of research.

This work is devoted to the study of cutoff regularization in the coordinate representation, which was first proposed in \cite{34} and successfully applied to a number of models \cite{Ivanov-Kharuk-2020,Ivanov-Kharuk-20222,Ivanov-Kharuk-2023,Iv-2024-1}. Since the first formulation, a set of important properties has been studied in \cite{Ivanov-2022,Iv-2024-2}. For example, a connection  with the theory of homogenization was obtained, a spectral representation was derived, and a criterion of applicability was obtained. Such properties make the proposed regularization very attractive and intuitive from a computational point of view.

The text discusses the three-loop renormalization of a five-dimensional scalar cubic model using the background field method and the regularization just mentioned. Despite the fact that the theory is super-renormalizable, the problem under study is new and relevant. It allows us to test the proposed regularization. In particular, it is shown that singular terms of the "non-local" type are successfully reduced, and the results are consistent with those previously obtained. The work also contains important calculations of a technical nature, which give new integral relations and are based on non-trivial reasoning and equalities. 

The structure of the work can be described as follows.  Section \ref{30:sec:prop} contains a problem statement with a description of the quantum action and regularization rules. In the final part of the section, the main results are formulated in the form of Theorem \ref{30-th-1}. Next, in Section \ref{30:sec:2-cor}, the first two quantum corrections are calculated, and Section \ref{30:sec:3-cor}, we study the third contribution and show an additional method for verifying the result. At the same time, calculations of a technical nature for finding asymptotic expansions for basic integrals are placed into Section \ref{30:sec:appen}. Section \ref{30:sec:next-cor}   examines a partial answer for the fourth correction. The conclusion contains useful comments and acknowledgements.

\section{Problem statement and results}
\label{30:sec:prop}

\subsection{The basic formulation}
Consider the standard 5-dimensional Euclidean space $\mathbb{R}^5$. Since the work uses a cutoff regularization in the coordinate representation, the dimension of the space remains fixed throughout all calculations (without deformation). The elements of $\mathbb{R}^5$ are notated by the letters $x$, $y$, $z$ (with possible indices), and their individual components are notated using the Greek indices $\mu$, $\nu$. Also, let the Latin indices $a$, $b$, $c$ take values in the set $\{1,...,n\}$, where $n\in\mathbb{N}$. Next, we define the scalar real field $\phi_a(\cdot)$ and the classical action $S[\,\cdot\,]$ for a cubic model in the general formulation
\begin{equation}\label{30-1}
S[\phi]=\int_{\mathbb{R}^5}\mathrm{d}^5x\,\bigg(t^a\phi_a(x)+\frac{1}{2}\phi_a(x)\Big(\delta^{ab}A_0(x)+M^{ab}\Big)\phi_b(x)+\frac{g^{abc}}{3!}\phi_a(x)\phi_b(x)\phi_c(x)\bigg),
\end{equation}
where $A_0(x)=-\partial_{x_\mu}\partial_{x^\mu}$ is the Laplace operator, and $t^a$, $M^{ab}$, $g^{abc}$ are completely symmetric coefficients. Also note that $g^{abc}$ plays the role of a coupling constant.

After the formulation of the classical theory, let us move on to the quantum action. It is defined by the following functional integral
\begin{equation}\label{30-2}
e^{-W/\hbar}=\int_{\mathcal{H}}\mathcal{D}\phi\, e^{-S[\phi]/\hbar},
\end{equation}
where $\hbar$ is the Planck constant, $\mathcal{H}$ is a functional space whose elements satisfy certain fixed boundary conditions. As is known, object (\ref{30-2}) contains divergences, which in the a perturbative approach are manifested by the presence of divergent integrals in the coefficients of the asymptotic series. This leads to the need to introduce a regularization, which in the framework of this work consists in adding a regularizing additive $S[\phi,\Lambda]$ to the classical action (\ref{30-1}). According to the previously proposed method, see Section 4 in \cite{Iv-2024-1}, the additive has the following form
\begin{equation*}\label{30-3}
S[\phi,\Lambda]=\frac{1}{2}\int_{\mathbb{R}^5}\mathrm{d}^5x\,\phi_a(x)\Big(A^\Lambda_0(x)-A_0^{\phantom{1}}(x)\Big)\phi_a(x),
\end{equation*} 
where $A^\Lambda_0(x)$ is a deformed Laplace operator such that the corresponding Green's function $R_0^\Lambda(\cdot)$ is obtained by the transition
\begin{equation}\label{30-4}
R_0^{\phantom{1}}(x)=\frac{1}{8\pi^2|x|^3}
\to R_0^\Lambda(x)=\frac{\Lambda^3}{8\pi^2}\mathbf{f}\big(|x|^2\Lambda^2\big)+\frac{1}{8\pi^2}
\begin{cases}
\,\,\,\Lambda^3, &|x|\leqslant 1/\Lambda;\\
|x|^{-3}, &|x|>1/\Lambda.
\end{cases}
\end{equation}
Here $\Lambda$ is a dimensional regularization parameter, and $\mathbf{f}(\cdot)$ is an auxiliary regularization function. The last one has the following properties
\begin{equation*}\label{30-5}
\mathbf{f}(\cdot)\in C\big([0,+\infty),\mathbb{R}\big),\,\,\, \text{supp}(\mathbf{f(\cdot)})\subset [0,1], \,\,\,A_0(x)\Lambda^3\mathbf{f}(|x-y|^2\Lambda^2)\xrightarrow{\Lambda\to +\infty} 0.
\end{equation*}
In particular, the latter property guarantees convergence of the deformed function $R^\Lambda_0(\cdot)$ to the original one $R_0^{\phantom{1}}(\cdot)$, as well as the combination $A_0^{\phantom{1}}(x)R_0^\Lambda(x-y)$ to $\delta(x-y)$ in the sense of generalized functions on the Schwartz class $S(\mathbb{R}^5)$. As a result, the regularized quantum action has the form
\begin{equation*}\label{30-6}
W\xrightarrow{\mbox{\footnotesize{reg.}}} W[\Lambda]=-\hbar\ln\bigg(\int_{\mathcal{H}}\mathcal{D}\phi\, e^{-(S[\phi]+S[\phi,\Lambda])/\hbar}\bigg)-\sum_{n=1}^{+\infty}\hbar^n\kappa_n,
\end{equation*}
where the numerical coefficients $\kappa_n$ shift a singular density independent of the boundary conditions. Such an object no longer contains divergences, although it has coefficients singular in the parameter $\Lambda$. In order to reduce them, it is necessary to apply the renormalization. Given that the cubic model in five-dimensional space is super-renormalizable, see \cite{105}, the procedure is multiplicative in nature and consists in shifting only two coefficients
\begin{equation}\label{30-7}
t^a\to t^a_r = t^a+\sum_{k=1}^{+\infty} \hbar^k t^a_{r,k},\,\,\,M^{ab}\to M^{ab}_r=M^{ab}+\sum_{k=1}^{+\infty}\hbar^kM^{ab}_{r,k}.
\end{equation}\label{30-8}
In this case, the renormalized actions are obtained as a result of transitions
\begin{equation*}
S[\phi]
\xrightarrow{\mbox{\footnotesize{ren.}}}
S_{\mathrm{ren}}[\phi]=S[\phi]\Big|_{t^a\to t^a_r,\,\, M^{ab}\to M^{ab}_r },
\end{equation*}
\begin{equation}\label{30-9}
W[\Lambda]
\xrightarrow{\mbox{\footnotesize{ren.}}} W_{\mathrm{ren}}[\Lambda]=-\hbar\ln\bigg(\int_\mathcal{H}\mathcal{D}\phi\,e^{-(S_{\mathrm{ren}}[\phi]+S_\Lambda[\phi])/\hbar}\bigg)-\sum_{n=1}^{+\infty}\hbar^n\kappa_n.
\end{equation}
The last object no longer contains singular terms, and the limit value $W_{\mathrm{ren}}[+\infty]$ is finite.

To find the renormalization coefficients (\ref{30-7}), it is convenient to decompose the quantum action (\ref{30-9}) into a series. To do this, we use the background field method \cite{102,103,23,24,25,26}, which consists in the following shift
\begin{equation*}\label{30-10}
\phi_a(x) \to B_a(x)+\sqrt{\hbar}\phi_a(x),
\end{equation*}
where $B_a(x)$ is a background field. It is a solution to a quantum equation of motion $Q_a[B](x)=0$ and satisfies the boundary conditions for the functions from $\mathcal{H}$. Therefore, the quantum action $W_{\mathrm{ren}}[\Lambda]$ is actually a functional $W_{\mathrm{ren}}[B,\Lambda]$, depending on the background field $B_a(x)$.

\begin{figure}[h]
	\center{\includegraphics[width=0.38\linewidth]{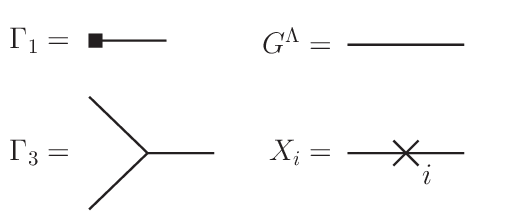}}
	\caption{Elements of the diagram technique.}
	\label{30-pic-1}
\end{figure}

Let us introduce a number of auxiliary definitions
\begin{equation*}\label{30-11}
\Gamma_1[\phi]=\int_{\mathbb{R}^5}\mathrm{d}^5x\bigg( t^a_r\phi_a(x)+B_a(x)A_0^\Lambda(x)\phi_a(x)+B_a(x)M^{ab}_r\phi_b(x)+\frac{g^{abc}}{2}B_a(x)B_b(x)\phi_c(x)\bigg),
\end{equation*}
\begin{equation*}\label{30-12}
\Gamma_3[\phi]=\int_{\mathbb{R}^5}\mathrm{d}^5x\,g^{abc}\phi_a(x)\phi_b(x)\phi_c(x),
\end{equation*}
\begin{equation*}\label{30-13}
S_a[\phi]=\int_{\mathbb{R}^5}\mathrm{d}^5x\,\phi_a(x),
\end{equation*}
\begin{equation*}\label{30-14}
S_{ab}[\phi]=\int_{\mathbb{R}^5}\mathrm{d}^5x\,\phi_a(x)\phi_b(x),\,\,\,\,\,\,X_i[\phi]=M^{ab}_{r,i}S_{ab}[\phi],
\end{equation*}
\begin{equation*}\label{30-15}
V^{ab}(x)=M^{ab}+g^{abc}B_c(x),
\end{equation*}
\begin{equation}\label{30-16}
\big(A_0^\Lambda(x)\delta^{ab}+V^{ab}(x)\big)G^\Lambda_{bc}(x,y)=\delta^{ac}\delta(x-y).
\end{equation}
We also define the elements of the diagram technique, which are shown in Fig. \ref{30-pic-1}.
In our case, the quantum equation of motion is represented by the sum of all strongly connected diagrams with one external free line, and the quantum action is equal to the sum of all strongly connected vacuum diagrams and is represented by the formula
\begin{multline}\label{30-17}
W_{\mathrm{ren}}[B,\Lambda]=S_{\mathrm{ren}}[B]-\bigg(\frac{\hbar}{2}\ln\det(G^\Lambda)+\hbar\kappa_1\bigg)-\\
-\Bigg[\hbar\exp\Bigg(-\frac{1}{2}\sum_{k=1}^{+\infty}\hbar^k X_k[\delta_j]-\frac{\hbar^{1/2}}{3!}\Gamma_3[\delta_j]\Bigg)e^{g[G^\Lambda,j]}\bigg|^{1\mathrm{PI}}_{j=0}+\sum_{n=2}^{+\infty}\hbar^n\kappa_n\Bigg],
\end{multline}
where $j_a(x)$ is an auxiliary field, $\delta_{j_a(x)}$ is the variational derivative with respect to the field $j_a(x)$, and
\begin{equation*}\label{30-18}
g[G^\Lambda,j]=\frac{1}{2}\int\int_{\mathbb{R}^5\times\mathbb{R}^5}\mathrm{d}^5x\mathrm{d}^5y\,j_a^{\phantom{1}}(x)G_{ab}^\Lambda(x,y)j_b^{\phantom{1}}(y).
\end{equation*}
Also, the symbol "$1\mathrm{PI}$" means that only strongly connected diagrams remain in total. Note that due to the quantum equation of motion, the vertex with one line $\Gamma_1[\,\cdot\,]$ do not appear in representation (\ref{30-17}).

\begin{figure}[h]
	\center{\includegraphics[width=0.34\linewidth]{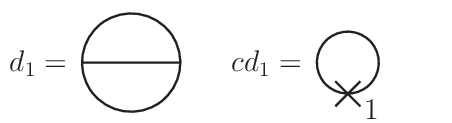}}
	\caption{The diagram $d_1$ and the counter diagram $cd_1$.}
	\label{30-pic-2}
\end{figure}

\begin{figure}[h]
	\center{\includegraphics[width=0.49\linewidth]{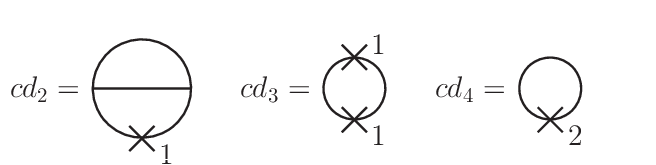}}
	\caption{The counter diagrams $cd_2$--$cd_4$.}
	\label{30-pic-3}
\end{figure}

\begin{figure}[h]
	\center{\includegraphics[width=0.34\linewidth]{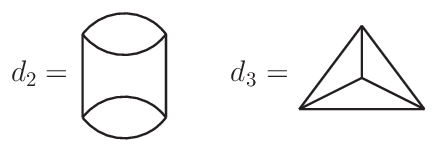}}
	\caption{The diagrams $d_2$ and $d_3$.}
	\label{30-pic-4}
\end{figure}

Considering that the renormalized action $W_{\mathrm{ren}}[B,\Lambda]$ does not contain singular terms, it can be argued that they are absent in each order of $\hbar^k$ with respect to the Planck constant. In this way, we obtain relations for the coefficients of the renormalization constants. In this paper, the first three ratios are investigated. They have the form
\begin{equation}\label{30-19}
t^a_{r,1}S_a[B]+\frac{1}{2}M^{ab}_{r,1}S_{ab}[B]\stackrel{\text{s.p.}}{=}\frac{1}{2}\ln\det(G^\Lambda)+\kappa_1,
\end{equation}
\begin{equation}\label{30-20}
t^a_{r,2}S_a[B]+\frac{1}{2}M^{ab}_{r,2}S_{ab}[B]\stackrel{\text{s.p.}}{=}\frac{1}{12}d_1-\frac{1}{2}cd_1+\kappa_2,
\end{equation}
\begin{equation}\label{30-21}
t^a_{r,3}S_a[B]+\frac{1}{2}M^{ab}_{r,3}S_{ab}[B]\stackrel{\text{s.p.}}{=}\frac{1}{16}d_2+\frac{1}{24}d_3-\frac{1}{4}cd_2+\frac{1}{4}cd_3-\frac{1}{2}cd_4+\kappa_3,
\end{equation}
where the sign $\stackrel{\text{s.p.}}{=}$ means equality of singular parts. We also use notations for the diagrams shown in Fig. \ref{30-pic-2}, \ref{30-pic-3}, and \ref{30-pic-4}.

Let us draw attention to one important property of the regularization used. It preserves the connection of the quantum action $W[B]$ and the quantum equation of motion $Q_a[B](x)$, see Section 4 in \cite{Iv-2024-1}, of the form
\begin{equation}\label{30-q-1}
\frac{\delta W[B]}{\delta B_a(x)}=Q_a[B](x)
\end{equation}
both after the regularization and after the renormalization. The relations themselves have the form
\begin{equation}\label{30-q-2}
\eqref{30-q-1}\xrightarrow{\mbox{\footnotesize{reg.}}}
\frac{\delta W[B,\Lambda]}{\delta B_a(x)}=Q_a[B,\Lambda](x)
\xrightarrow{\mbox{\footnotesize{ren.}}}
\frac{\delta W_{\mathrm{ren}}[B,\Lambda]}{\delta B_a(x)}=
Q^{\mathrm{ren}}_{a}[B,\Lambda](x).
\end{equation}
The preservation of this type of connections ensures the renormalization of the quantum equation of motion. In other words, found coefficients of the renormalization constants remove singularities not only in the quantum action, but also in the quantum equation of motion. In order to verify the validity of the transitions from \eqref{30-q-2}, it is enough to check that the relations for the variational derivatives of all elements of the diagram technique retain their form after regularization and renormalization. Given the fact that the vertices are not deformed, it is sufficient to consider only the Green's function. Equation \eqref{30-16} implies the relation
\begin{equation*}\label{30-q-3}
\frac{\delta}{\delta B_c(z)}G_{ab}(x,y)=-
G_{ad}(x,z)g^{dce}G_{eb}(z,y).
\end{equation*}
Using the considerations from \cite{Iv-2024-1} and the explicit form of decomposition by powers of the potential, one can make sure that the following equality holds
\begin{equation*}\label{30-q-4}
\frac{\delta}{\delta B_c(z)}G_{ab}^\Lambda(x,y)=-
G_{ad}^\Lambda(x,z)g^{dce}G_{eb}^\Lambda(z,y),
\end{equation*}
which guarantees the feasibility of the declared properties.

\subsection{Results}

Let us define a number of auxiliary coefficients
\begin{equation*}\label{30-d-1}
h_1^{a_1a_2}=g^{a_1c_1c_2}g^{a_2c_1c_2},\,\,\,
g_1^{a_1a_2a_3}=g^{a_1c_1c_2}g^{a_2c_2c_3}g^{a_3c_3c_1},
\end{equation*}
\begin{equation*}\label{30-d-2}
h^{a_1a_2}_2=h_1^{c_1c_2}g^{c_1c_3a_1}g^{c_2c_3a_2},\,\,\,
h^{a_1a_2}_3=g_1^{a_1c_1c_2}g^{c_1c_2a_2}.
\end{equation*}

\begin{theorem}\label{30-th-1}
Taking into account all the above, the singular terms for the first three quantum corrections are proportional to the linear combination of the functionals $S_{a}[B]$ and $S_{ab}[B]$. The values for the coefficients of the renormalization constants \eqref{30-7}, satisfying the relations \eqref{30-19}--\eqref{30-21}, have the form
\begin{center}
\begin{tabular}{cc}
$t_{r,1}^a\to$\mbox{~see~\eqref{30-29},}&$M_{r,1}^{ab}\to$\mbox{~see~\eqref{30-30},}\\
$t_{r,2}^a\to$\mbox{~see~\eqref{30-40},}&$M_{r,2}^{ab}\to$\mbox{~see~\eqref{30-42},}\\
$t_{r,3}^a\to$\mbox{~see~\eqref{30-79},}&$M_{r,3}^{ab}\to$\mbox{~see~\eqref{30-82}.}
\end{tabular}
\end{center}
The partial answer for $t_{r,4}^a$ is given in formula \eqref{30-95}. Also $t_{r,i+1}^a\stackrel{\text{s.p.}}{=}0$ and $M_{r,i}^{ab}\stackrel{\text{s.p.}}{=}0$ for $i\geqslant3$.
\end{theorem}

Additionally, we write out a special case, when $n=1$, $M^{11}=m^2$ and summation in classical action (\ref{30-1}) is absent. For convenience, we present all coefficients without indices, since they take only one value.
Using the above answers, for the simplified model the equalities are valid
\begin{align*}
	t_r=t+&\hbar\bigg(-\frac{\Lambda^3g}{2}R^1_0(0)+\frac{m^2g}{2}\Lambda\alpha_0(\mathbf{f})+\tilde{t}_{r,1}\bigg)
	+\\
	+&\hbar^2\bigg(-\frac{g^3\Lambda^2}{4}\Big(\alpha_1(\mathbf{f})-\alpha_0^2(\mathbf{f})\Big)+\frac{g\Lambda}{2}\alpha_0(\mathbf{f})\tilde{M}_{r,1}+\frac{Lg^3m^2}{12(16\pi^2)^2}+\tilde{t}_{r,2}\bigg)+
	\\
	+&\hbar^3\bigg(\frac{\Lambda Lg^5}{24(16\pi^2)^2}\alpha_0(\mathbf{f})-\frac{\Lambda g^5}{8}\Big(\alpha_2(\mathbf{f})+2\alpha_3(\mathbf{f})+2\alpha_4(\mathbf{f})\Big)+\frac{\Lambda g}{2}\alpha_0(\mathbf{f})\tilde{M}_{r,2}+
	\\
	&~~~~~~~~~~~~~~~~~~~~~~~~~~~~
	~~~~~~~~~~~~~~~~~~~~~~~~~~~~
	+\frac{Lg^3}{12(16\pi^2)^2}\tilde{M}_{r,1}+\tilde{t}_{r,3}\bigg)+\mathcal{O}\big(\hbar^4\big),
	\notag
\end{align*}
\begin{equation}\label{30-87}
	M_r=m^2+\hbar\bigg(\frac{\Lambda g^2}{2}\alpha_0(\mathbf{f})+\tilde{M}_{r,1}\bigg)+\hbar^2\bigg(\frac{Lg^4}{12(16\pi^2)^2}+\tilde{M}_{r,2}\bigg)+\hbar^3\tilde{M}_{r,3}+\mathcal{O}\big(\hbar^4\big),
\end{equation}
where $L=\ln(\Lambda/\sigma)$. Note that the result is consistent with the previously obtained, see \cite{34}.

Immediately note that the calculation of $\alpha_1(\mathbf{f})$--$\alpha_4(\mathbf{f})$ for a specific function $\mathbf{f}(\cdot)$ is a less important problem since such numbers are included only in the linear term. At the same time, their calculation is not trivial and requires the use of some numerical methods.

Consider $\alpha_0(\mathbf{f})$. It is included in the renormalization of the mass, see \eqref{30-30} and \eqref{30-87}, and is represented by formula \eqref{30-27}. Using the substitution of an explicit type of deformation, we get
\begin{equation*}\label{30-f-1}
\alpha_0(\mathbf{f})=\int_{\mathbb{R}^5}\mathrm{d}^5x\,\Big(R^1_0(x)\Big)^2
=\frac{1}{3(8\pi^2)}\bigg(
1+\int_0^1\mathrm{d}s\,s^4\Big(\mathbf{f}\big(s^2\big)+1\Big)^2
\bigg)\geqslant\frac{1}{3(8\pi^2)}.
\end{equation*}
In this case, it is interesting to pay attention to two specific choices of the function $\mathbf{f}(\cdot)$.
\begin{itemize}
	\item  The first one is defined by the equality $\mathbf{f}_1\big(s^2\big)=0$ and is the most trivial. In this case, $\alpha_0(\mathbf{f}_1)=1/(20\pi^2)$. However, such a choice does not satisfy the applicability condition for cutoff in the coordinate representation, see \cite{Iv-2024-2}.
	\item The second choice is determined by the equality $\mathbf{f}_2\big(s^2\big)=7-9s+2s^3$. According to the results from \cite{Iv-2024-2}, it satisfies the criterion of applicability. In this case, we have $\alpha_0(\mathbf{f}_2)=167/(2310\pi^2)>\alpha_0(\mathbf{f}_1)$.
\end{itemize}

\section{The first two corrections}
\label{30:sec:2-cor}

Let us consider the asymptotic expansion of the Green's function $G^\Lambda_{ab}(x,y)$ near the diagonal $x\sim y$. According to the works \cite{29,34,Iv-2024-1,30-1-1}, the series has the following form
\begin{multline}\label{30-22}
G^\Lambda_{ab}(x,y)=R_0^\Lambda(x-y)\delta_{ab}-R_1^\Lambda(x-y)\frac{V^{ab}(x)+V^{ab}(y)}{2}+\\+R^\Lambda_2(x-y)\frac{V^{ac}(x)V^{cb}(x)+V^{ac}(y)V^{cb}(y)}{4}+PS_{1ab}^\Lambda(x,y),
\end{multline}
where the deformed function $R^\Lambda_0(\cdot)$ was defined in (\ref{30-4}), and the rest ones are defined by the equalities
\begin{equation}\label{30-23}
R_1^\Lambda(z)=\int_{\mathrm{B}_{1/\sigma}}\mathrm{d}^5x\,R_0^\Lambda(x)R_0^\Lambda(x+z),
\end{equation}  
\begin{equation}\label{30-24}
R_2^\Lambda(z)=2\int\int_{\mathrm{B}_{1/\sigma}\times \mathrm{B}_{1/\sigma}}\mathrm{d}^5x\mathrm{d}^5y\,R_0^\Lambda(x)R_0^\Lambda(x+y)R_0^\Lambda(y+z)-2\tilde{c}_2,
\end{equation}
\begin{equation}\label{30-25}
\tilde{c}_2=\int\int_{\mathrm{B}_{1/\sigma}\times\mathrm{B}_{1/\sigma}}\mathrm{d}^5x\mathrm{d}^5y\,R_0(x)R_0(x+y)R_0(y).
\end{equation}
In the latest formulas, $\mathrm{B}_{1/\sigma}$ is a closed ball with the center the origin and the radius of $1/\sigma$. The parameter $\sigma$ is auxiliary. It is dimensional, finite, and fixed. The Green's function does not depend on it. Decomposition (\ref{30-22}) is specially written out in a symmetrical form for convenience. The function $PS_1^\Lambda(x,y)$ contains residual terms, which, among other things, ensure the feasibility of boundary conditions. It is shown below that such function does not appear in contributions with singular coefficients.

Let us move on to calculating the first loop. The right part of formula (\ref{30-19}) is expressed as a series depending on the Green's function. Using decomposition (\ref{30-22}), the formula from \cite{Iv-2024-1,29-hp,29-3} can be represented as
\begin{multline}\label{30-26}
\frac{1}{2}\ln\det\Big(G^\Lambda/G^\Lambda|_{B=0}\Big)\stackrel{\text{s.p.}}{=}-\frac{1}{2}R_0^\Lambda(0)\int_{\mathbb{R}^5}\mathrm{d}^5x\,\Big(V^{aa}(x)-M^{aa}\Big)+\\+\frac{1}{4}\Bigg(\int_{\mathrm{B}_{1/\sigma}}\mathrm{d}^5y\,\Big(R_0^\Lambda(y)\Big)^2\Bigg)\int_{\mathbb{R}^5}\mathrm{d}^5x\,\Big(V^{ab}(x)V^{ba}(x)-M^{ab}M^{ba}\Big).
\end{multline}
Let us introduce the following notation for convenience
\begin{equation}\label{30-27}
\mathrm{A}(\sigma)=\int_{\mathrm{B}_{1/\sigma}}\mathrm{d}^5x\, \Big(R^\Lambda_0(x)\Big)^2=\Lambda\alpha_0(\mathbf{f})-\frac{\sigma}{3(8\pi^2)},
\end{equation}
then the right hand part of (\ref{30-26}) contains a singular contribution of the form
\begin{equation*}\label{30-28}
S_a[B]\bigg(-\frac{1}{2}R_0^\Lambda(0)g^{abb}+\frac{1}{2}\mathrm{A}(\sigma)M^{bc}g^{abc}\bigg)+S_{ab}[B]\bigg(\frac{1}{4}\mathrm{A}(\sigma)g^{acd}g^{bcd}\bigg).
\end{equation*}
Thus, the constants from the left hand side of (\ref{30-19}) have the following representations
\begin{equation}\label{30-29}
t^a_{r,1}=-\frac{\Lambda^3}{2}R_0^1(0)g^{abb}+\frac{\Lambda}{2}\alpha_0(\mathbf{f})M^{bc}g^{abc}+\tilde{t}^{a}_{r,1},
\end{equation}
\begin{equation}\label{30-30}
M^{ab}_{r,1}=\frac{\Lambda}{2}\alpha_0(\mathbf{f})h_1^{ab}+\tilde{M}^{ab}_{r,1},
\end{equation}
where $\tilde{t}^a_{r,1}$ and $\tilde{M}^{ab}_{r,1}$ are symmetric constants. They are auxiliary in nature, do not contain singularities, and play the role of "boundary conditions" in the renormalization procedure. Note that both coefficients have power-law behavior at $\Lambda\to+\infty$.

Let us move on to the study of the second correction. In this case, in addition to the diagram $d_1$, there is also a counter diagram $cd_1$. The method of calculating them has been shown many times before, see \cite{34,Iv-2024-1}, so we present only the final result for our regularization case. For the diagram $d_1$, the expression looks like
\begin{align}
\nonumber
-&3\mathrm{I}_1(\Lambda,\sigma)S_a[B]g_1^{abb}\\\nonumber
+&3\mathrm{I}_2(\Lambda,\sigma)\Big(2S_a[B]g^{ab_1b_2}_1M^{b_1b_2}+S_{a_1a_2}[B]h_3^{a_1a_2}\Big)\\\nonumber
+&\frac{3}{2}\mathrm{I}_3(\Lambda,\sigma)\Big(2S_a[B]g^{ab_1b_2}_{\phantom{1}}M^{b_2b_3}_{\phantom{1}}h_1^{b_3b_1}+S_{a_1a_2}[B]h_2^{a_1a_2}\Big)\\
\label{30-34}
+&3\mathrm{A}(\sigma)\int_{\mathbb{R}^5}\mathrm{d}^5x\,PS^\Lambda_{1a_1a_2}(x,x)h_1^{a_1a_2},
\end{align}
where the notations for auxiliary integrals was introduced
\begin{equation}\label{30-35}
\mathrm{I}_1(\Lambda,\sigma)=\int_{\mathrm{B}_{1/\sigma}}\mathrm{d}^5x\,\Big(R^\Lambda_0(x)\Big)^2R^\Lambda_1(x),
\end{equation}
\begin{equation}\label{30-36}
\mathrm{I}_2(\Lambda,\sigma)=\int_{\mathrm{B}_{1/\sigma}}\mathrm{d}^5x\,\Big(R^\Lambda_1(x)\Big)^2R^\Lambda_0(x),
\end{equation}
\begin{equation}\label{30-37}
\mathrm{I}_3(\Lambda,\sigma)=\int_{\mathrm{B}_{1/\sigma}}\mathrm{d}^5x\,\Big(R^\Lambda_0(x)\Big)^2R^\Lambda_2(x).
\end{equation}

Further, given the fact that $R_2^\Lambda(0)$ behaves like $\Lambda^{-1}$, see (\ref{30-p42}), the answer for the singular part of the counter diagram takes the form
\begin{equation}
-\frac{1}{2}\mathrm{A}(0)\mathrm{A}(\sigma)S_a[B]g_1^{abb}-\mathrm{A}(\sigma)S_a[B]\tilde{M}^{b_1b_2}_{r,1}g^{b_1b_2a}+
\frac{1}{2}\mathrm{A}(0)\int_{\mathbb{R}^5}d^5x\, PS_{1a_1a_2}^\Lambda(x,x)h_1^{a_1a_2}\label{30-39}.
\end{equation}
For the final result, consider the linear combination $d_1/12-cd_1/2$. Note that the non-local parts (\ref{30-34}) and (\ref{30-39}) are reduced. The local parts contribute to the coefficients from the left hand side of (\ref{30-20}). Taking into account the asymptotic expansions (\ref{30-p7}), (\ref{30-p8}), and (\ref{30-p14}), the answer is written as follows
\begin{align}
t^a_{r,2}=&-\frac{\Lambda^2}{4}\Big(\alpha_1(\mathbf{f})-\alpha_0^2(\mathbf{f})\Big)g_1^{abb}+\frac{\Lambda}{2}\alpha_0(\mathbf{f})\tilde{M}^{b_1b_2}_{r,1}g^{b_1b_2a}+\label{30-40}\\
&+\frac{L}{12(16\pi^2)^2}\Big(2g^{ab_1b_2}_1M^{b_1b_2}-g^{ab_1b_2}_{\phantom{1}}M^{b_2b_3}_{\phantom{1}}h_1^{b_3b_1}\Big)+\tilde{t}^a_{r,2},\nonumber
\end{align}
\begin{equation}\label{30-42}
M^{a_1a_2}_{r,2}=\frac{L}{12(16\pi^2)^2}\Big(2h_3^{a_1a_2}-h_2^{a_1a_2}\Big)+\tilde{M}^{a_1a_2}_{r,2},
\end{equation}
where $\tilde{t}^a_{r,2}$ and $\tilde{M}^{a_1a_2}_{r,2}$ are additional finite symmetric coefficients.

\section{The third correction}
\label{30:sec:3-cor}
Let us start our consideration of the third correction with the diagram $d_2$. To do this, it is convenient to introduce an auxiliary integral
\begin{equation}\label{30-43}
\rho^{ab}_1(x,y)=\int_{\mathbb{R}^5}\mathrm{d}^5z\,g^{aa_1a_2}G^\Lambda_{a_1b_1}(x,z)G^\Lambda_{a_2b_2}(x,z)g^{b_1b_2b_3}G^\Lambda_{b_3b}(z,y).
\end{equation}
It is clear that when the regularization is removed, such integral diverges, since the first two multipliers on the right hand side are proportional to $|x-z|^{-6}$ for $x\sim z$. 
Using this notation, the diagram $d_2$ has the form
\begin{equation}\label{30-44}
d_2=\int_{\mathbb{R}^5\times\mathbb{R}^5}\mathrm{d}^5x\mathrm{d}^5y\,\rho^{ab}_1(x,y)\rho^{ba}_1(y,x).
\end{equation}
To analyze such diagram, we use method based on adding and subtracting, which was analyzed in detail for the quartic interaction in \cite{Iv-2024-1}.

Let us define an additional auxiliary function
\begin{equation}\label{30-45}
\rho_2^{ab}(x,y)=\Big(\mathrm{A}(0)h_1^{aa_3}+2\tilde{M}^{aa_3}_{r,1}\Big)G^\Lambda_{a_3b}(x,y),
\end{equation}
then representation (\ref{30-44}) for the diagram $d_2$ takes the form
\begin{align*}
2\int_{\mathbb{R}^5\times\mathbb{R}^5}\mathrm{d}^5x\mathrm{d}^5y\,\rho_1^{ab}(x,y)&\rho_2^{ba}(y,x)-\int_{\mathbb{R}^5\times\mathbb{R}^5}\mathrm{d}^5x\mathrm{d}^5y\,\rho_2^{ab}(x,y)\rho_2^{ba}(y,x)+
\\
&+\int_{\mathbb{R}^5\times\mathbb{R}^5}\mathrm{d}^5x\mathrm{d}^5y\,\Big(\rho_1^{ab}(x,y)-\rho_2^{ab}(x,y)\Big)\Big(\rho_1^{ba}(y,x)-\rho_2^{ba}(y,x)\Big).
\end{align*}
This representation is notable in that the first two terms in it are actually the counter diagrams $4cd_2$ and $-4cd_3$. It is easy to verify this by substituting the functions from (\ref{30-43}) and (\ref{30-45}). Consequently, the following relation has been proved
\begin{equation}\label{30-48}
\frac{d_2}{16}-\frac{cd_2}{4}+\frac{cd_3}{4}+\tilde{\kappa}_1=\frac{1}{16}\int_{\mathbb{R}^5\times\mathbb{R}^5}\mathrm{d}^5x\mathrm{d}^5y\,\Big(\rho_1^{ab}(x,y)-\rho_2^{ab}(x,y)\Big)\Big(\rho_1^{ba}(y,x)-\rho_2^{ba}(y,x)\Big)+\tilde{\kappa}_2,
\end{equation}
where $\tilde{\kappa}_1$ and $\tilde{\kappa}_2$ subtract divergences independent of the background field. It is this linear combination that appears on the right hand side of (\ref{30-21}). Therefore, the analysis of the three diagrams is reduced to the study of integral (\ref{30-48}).

Let us introduce the following transformation for convenience
\begin{equation}\label{30-49}
\hat{R}^\Lambda_i(x)=\int_{\mathbb{R}^5}\mathrm{d}^5y\,\Big(R^\Lambda_0(y)\Big)^2R^\Lambda_i(y+x)-\mathrm{A}(0)R^\Lambda_i(x).
\end{equation}
The peculiarity of such transformation is that the transformed function exists when the regularization is removed, since the divergent contribution is subtracted. Moreover, $\hat{R}^\Lambda_0(x)$ has the behavior of $|x|^{-4}$, and $\hat{R}^\Lambda_1(x)\sim |x|^{-2}$. Also note that in the right part of (\ref{30-48}), singular terms appear only if the density has the behavior $|x-y|^{-5}$ or worse. Therefore, the following transition is valid
\begin{equation*}\label{30-50}
\rho_1^{ab}(x,y)-\rho_2^{ab}(x,y)\to\sum_{i=3}^8\rho_i^{ab}(x,y),
\end{equation*}
where
\begin{equation}\label{30-51}
\rho_3^{ab}(x,y)=\int_{\mathbb{R}^5}\mathrm{d}^5z\,h_1^{ab}\Big(R^\Lambda_0(x-z)\Big)^2R^\Lambda_0(z-y)-\mathrm{A}(0)h_1^{ab}R^\Lambda_0(x-y)=\hat{R}^\Lambda_0(x-y)h_1^{ab},
\end{equation}
\begin{equation}\label{30-52}
\rho_4^{ab}(x,y)=-2\tilde{M}^{ab}_{r,1}R^\Lambda_0(x-y),
\end{equation}
\begin{equation}\label{30-53}
\rho_5^{ab}(x,y)=-\hat{R}^\Lambda_1(x-y)h_1^{ab_3}g^{b_3b_4b}B_{b_4}(x),
\end{equation}
\begin{equation}\label{30-54}
\rho_6^{ab}(x,y)=2R^\Lambda_1(x-y)\tilde{M}^{aa_3}_{r,1}g^{a_3a_4b}B_{a_4}(x),
\end{equation}
\begin{equation}\label{30-55}
\rho_7^{ab}(x,y)=-2\int_{\mathbb{R}^5}\mathrm{d}^5z\, R^\Lambda_0(x-z)R_1^\Lambda(x-z)R^\Lambda_0(z-y)g^{aa_1a_2}g^{a_1a_4a_5}B_{a_4}(x)g^{a_5a_2b},
\end{equation}
\begin{equation}\label{30-56}
\rho_8^{ab}(x,y)=2\int_{\mathbb{R}^5}\mathrm{d}^5z\,R^\Lambda_0(x-z)R_0^\Lambda(z-y)g^{aa_1a_2}PS^\Lambda_{1a_1a_3}(x,x)g^{a_2a_3b}.
\end{equation}
Using the latter definitions, the singular part of the linear combination from (\ref{30-48}), depending on the background field, can be represented as follows
\begin{equation}\label{30-57}
\frac{d_2}{16}-\frac{cd_2}{4}+\frac{cd_3}{4}+\tilde{\kappa}_1\stackrel{\text{s.p.}}{=}\frac{2}{16}\sum_{i=5}^8J_{3i}[B]+\frac{2}{16}\sum_{i=5,7}J_{4i}[B],
\end{equation} 
where we have used the symmetry of $\rho^{ab}_3(x,y)$ and $\rho^{ab}_4(x,y)$ by indices and arguments, and the notation for $i,k\in\{3,...,8\}$
\begin{equation*}\label{30-58}
J_{ik}[B]=\int_{\mathrm{B}_{1/\sigma}\times\mathbb{R}^5}\mathrm{d}^5x\mathrm{d}^5y\,\rho^{ba}_i(y,x+y)\rho^{ba}_k(y,x+y).
\end{equation*}
In addition, for convenience, we define a number of singular integrals
\begin{equation}\label{30-60}
\mathrm{I}_4(\Lambda,\sigma)=\int_{\mathrm{B}_{1/\sigma}}\mathrm{d}^5x\,\hat{R}_0^\Lambda(x)\hat{R}^\Lambda_1(x),
\end{equation}
\begin{equation}\label{30-61}
\mathrm{I}_5(\Lambda,\sigma)=\int_{\mathrm{B}_{1/\sigma}}\mathrm{d}^5x\,\hat{R}_0^\Lambda(x)R^\Lambda_1(x),
\end{equation}
\begin{equation}\label{30-62}
\mathrm{I}_6(\Lambda,\sigma)=\int_{\mathrm{B}_{1/\sigma}\times\mathbb{R}^5}\mathrm{d}^5x\mathrm{d}^5z\,\hat{R}_0^\Lambda(x)R^\Lambda_0(x-z)R^\Lambda_1(x-z)R^\Lambda_0(z),
\end{equation}
\begin{equation}\label{30-63}
\mathrm{I}_7(\Lambda,\sigma)=\int_{\mathrm{B}_{1/\sigma}\times\mathbb{R}^5}\mathrm{d}^5x\mathrm{d}^5z\,\hat{R}_0^\Lambda(x)R^\Lambda_0(x-z)R^\Lambda_0(z),
\end{equation}
\begin{equation}\label{30-64}
\mathrm{I}_8(\Lambda,\sigma)=\int_{\mathrm{B}_{1/\sigma}}\mathrm{d}^5x\,R_0^\Lambda(x)\hat{R}^\Lambda_1(x),
\end{equation}
\begin{equation}\label{30-65}
\mathrm{I}_9(\Lambda,\sigma)=\int_{\mathrm{B}_{1/\sigma}\times\mathbb{R}^5}\mathrm{d}^5x\mathrm{d}^5z\,R_0^\Lambda(x)R^\Lambda_0(x-z)R^\Lambda_1(x-z)R_0^\Lambda(z).
\end{equation}
Then the individual parts of formula (\ref{30-57}) can be rewritten as
\begin{equation}\label{30-66}
J_{35}[B]\stackrel{\text{s.p.}}{=}-\Lambda S_a[B]h_1^{a_1a_2}h_1^{a_2a_3}g^{a_1a_3a}\alpha_2(\mathbf{f}),
\end{equation}
\begin{equation}\label{30-67}
J_{36}[B]\stackrel{\text{s.p.}}{=}-\frac{L}{3(16\pi^2)^2} S_a[B]h_1^{a_1a_2}\tilde{M}^{a_2a_3}_{r,1} g^{a_1a_3a}_{\phantom{1}},
\end{equation}
\begin{equation}\label{30-68}
J_{37}[B]\stackrel{\text{s.p.}}{=}-2\bigg(\Lambda\alpha_3(\mathbf{f})-\frac{L\big(\mathrm{A}(\sigma)-\mathrm{A}(0)\big)}{6(16\pi^2)^2}\bigg) S_a[B]g^{ab_1b_2}_1h_1^{b_1b_2},
\end{equation}
\begin{equation}\label{30-69}
J_{38}[B]\stackrel{\text{s.p.}}{=}-\frac{L}{3(16\pi^2)^2} \int_{\mathbb{R}^5}\mathrm{d}^5x\,PS_{1ab}^\Lambda(x,x)g_{\phantom{1}}^{ac_1c_3}g_{\phantom{1}}^{bc_1c_2}h_1^{c_2c_3},
\end{equation}
\begin{equation}\label{30-70}
J_{45}[B]\stackrel{\text{s.p.}}{=}-\frac{L}{3(16\pi^2)^2} S_a[B]\tilde{M}^{a_1a_2}_{r,1}h_1^{a_2a_3} g^{a_1a_3a},
\end{equation}
\begin{equation}\label{30-71}
J_{47}[B]\stackrel{\text{s.p.}}{=}\frac{4L}{3(16\pi^2)^2} S_a[B]\tilde{M}^{a_1a_2}_{r,1}g^{a_1a_2a}_1,
\end{equation}
where we have used equalities (\ref{30-p17}), (\ref{30-p24}), (\ref{30-p27}), and (\ref{30-p31}). 

As a result, it remains to consider the counter diagram $cd_4$ and the diagram $d_3$. Let us start with the first one. Using answer (\ref{30-42}) for the second coefficient and the calculation scheme for $cd_1$, we write out the singular component, depending on the background field, in the form
\begin{align}
-\frac{1}{2}cd_4+\tilde{\kappa}_3=~\mathrm{A}(\sigma)S_a[B]&\bigg(\frac{L}{12(16\pi^2)^2}h_3^{a_1a_2}g^{a_1a_2a}-\frac{L}{24(16\pi^2)^2}h_2^{a_1a_2}g^{a_1a_2a}+\frac{1}{2}\tilde{M}^{a_1a_2}_{r,2}g^{a_1a_2a}\bigg)\label{30-72}\\
+&\bigg(\int_{\mathbb{R}^5}\mathrm{d}^5x\,PS^\Lambda_{1ab}(x,x)\bigg)\bigg(-\frac{L}{12(16\pi^2)^2}h^{ab}_3+\frac{L}{24(16\pi^2)^2}h^{ab}_2\bigg),\label{30-73}
\end{align}
where $\tilde{\kappa}_3$ subtracts a singular density independent of the background field. Let us move on to the last diagram $d_3$. In fact, it contains only two singular contributions, depending on the background field. This is due to the fact that the integral of the form
\begin{equation*}\label{30-74}
\int_{\mathbb{R}^5\times\mathbb{R}^5}\mathrm{d}^5x\mathrm{d}^5y\,R_0(x)R_0(y)R_0(x-y)R_0(x-z)R_0(y-z)
\end{equation*}
converges and is proportional to $|z|^{-5}$. At the same time, any replacement of $R_0(\cdot)$ with a smoother function will only weaken the behavior at zero with respect to the variable $z$. Therefore, in the diagram $d_3$, it is necessary to save five functions $R^\Lambda_0(\cdot)$, and choose the sixth one either equal to $R_1^\Lambda(\cdot)$, or the unit function. Then we get
\begin{equation}\label{30-75}
\frac{1}{24}d_3+\tilde{\kappa}_4=-\frac{1}{4}\mathrm{I}_{10}(\Lambda,\sigma)S_a[B]g^{aa_1a_2}h_3^{a_1a_2}+\frac{1}{4}\mathrm{I}_{11}(\Lambda\sigma)\int_{\mathbb{R}^5}\mathrm{d}^5x\,PS^\Lambda_{1a_1a_2}(x,x)h^{a_1a_2}_3,
\end{equation}
where
\begin{equation}\label{30-76}
\mathrm{I}_{10}(\Lambda,\sigma)=\int_{\mathrm{B}_{1/\sigma}\times\mathbb{R}^5\times\mathbb{R}^5}\mathrm{d}^5z\mathrm{d}^5x\mathrm{d}^5y\,R^\Lambda_1(z)R^\Lambda_0(z-y)R^\Lambda_0(z-x)R^\Lambda_0(x-y)R^\Lambda_0(x)R^\Lambda_0(y),
\end{equation}
\begin{equation}\label{30-77}
\mathrm{I}_{11}(\Lambda,\sigma)=\int_{\mathrm{B}_{1/\sigma}\times\mathbb{R}^5\times\mathbb{R}^5}\mathrm{d}^5z\mathrm{d}^5x\mathrm{d}^5y\,R^\Lambda_0(z-y)R^\Lambda_0(z-x)R^\Lambda_0(x-y)R^\Lambda_0(x)R^\Lambda_0(y),
\end{equation}
and $\tilde{\kappa}_4$ subtracts a singular density independent of the background field. Using the calculated singular terms for the last integrals, see (\ref{30-p37}) and (\ref{30-p39}), relation (\ref{30-75}) is rewritten as follows
\begin{equation}\label{30-78}
\frac{1}{24}d_3+\tilde{\kappa}_4=\Bigg(-\frac{\Lambda\alpha_4(\mathbf{f})}{4}-\frac{L\big(\mathrm{A}(\sigma)-\mathrm{A}(0)\big)}{12(16\pi^2)^2}\Bigg)S_a[B]g^{aa_1a_2}h_3^{a_1a_2}+\frac{L}{12(16\pi^2)^2}\int_{\mathbb{R}^5}\mathrm{d}^5x\,PS^\Lambda_{1a_1a_2}(x,x)h^{a_1a_2}_3.
\end{equation}
Consequently, taking into account answers (\ref{30-66})--(\ref{30-71}), in relations (\ref{30-48}), (\ref{30-72})--(\ref{30-73}), and (\ref{30-78}), the results for singular contributions are presented for separate diagrams and their linear combinations. We immediately note that the non-local part $PS^\Lambda_1$ is reduced in total, as well as a logarithmic divergence, which depends according to a power law on the auxiliary parameter $\sigma$, that is, $L\mathrm{A}(\sigma)$. Next, note that the mass term does not require renormalization. This result fully corresponds to the expectations from the general theory. Summing up the studied contributions, we get
\begin{align}
t^a_{r,3}&=\frac{\Lambda L\alpha_0(\mathbf{f})}{24(16\pi^2)^2}\Big(2g^{aa_1a_2}h^{a_1a_2}_3-g^{aa_1a_2}h^{a_1a_2}_2\Big)-\label{30-79}\\
&-\Lambda\bigg(\frac{\alpha_2(\mathbf{f})}{8}h^{a_1a_2}_1h^{a_2a_3}_1g^{a_1a_3a}+\frac{\alpha_3(\mathbf{f})}{4}h^{a_1a_2}_2g^{a_1a_2a}+\frac{\alpha_4(\mathbf{f})}{4}h^{a_1a_2}_3g^{a_1a_2a}-\frac{\alpha_0(\mathbf{f})}{2}\tilde{M}^{a_1a_2}_{r,2}g^{a_1a_2a}\bigg)\nonumber
\\
&-\frac{L}{24(16\pi^2)^2}\Big(2h^{a_1a_2}_1\tilde{M}^{a_2a_3}_{r,1}g^{a_1a_3a}-4\tilde{M}^{a_1a_2}_{r,1}g^{a_1a_2a}_1\Big)+\tilde{t}^a_{r,3},\nonumber
\end{align}
\begin{equation}\label{30-82}
M^{ab}_{r,3}=\tilde{M}^{ab}_{r,3},
\end{equation}
where $\tilde{t}^a_{r,3}$ and $\tilde{M}^{ab}_{r,3}$ are constants. Since renormalization of the mass term is not required in the third loop, it is reasonable to choose $\tilde{M}^{ab}_{r,3}=0$, as well as the subsequent coefficients in the higher loops.

It is worth saying a few words about the additional verification. It is known that divergences of the form $\Lambda^k L^n$ and $L^{n+1}$, where $k>0$ and $n>0$, can be calculated using only counter diagrams. Indeed, applying the operator $-\sigma\partial_{\sigma}$ to both sides of relations \eqref{30-19}--\eqref{30-21}, we obtain equalities for singular parts. At the same time, all diagrams on the right hand side  vanish, since they do not depend on the parameter $\sigma$, but only counter diagrams, containing renormalization coefficients with the parameter $\sigma$, remain.

In the case of the model under study, the above divergences are contained only in the third loop, see (\ref{30-79}). Therefore, differentiating equality (\ref{30-21}) leads to the following nontrivial relation
\begin{equation*}\label{30-83}
	-\sigma\frac{d}{d\sigma}\bigg(t^a_{r,3}S_a[B]+\frac{1}{2}M^{ab}_{r,3}S_{ab}[B]\bigg)\stackrel{\text{s.p.}}{=}-\sigma\frac{d}{d\sigma}\bigg(-\frac{1}{2}cd_4+\bar{\kappa}_3\bigg),
\end{equation*}
which in the right part contains the derivative of one counter diagram. Thus, such check in the case of a five-dimensional cubic model is uninformative, since it moves on to the differentiation of one counter diagram, see (\ref{30-72}) and (\ref{30-73}), which gives a direct contribution to the linear part, see (\ref{30-79}), proportional to the combination $\Lambda L$.

\section{On the fourth loop}
\label{30:sec:next-cor}

Using the example of the five-dimensional cubic model, it was shown that the renormalization constant for the mass coefficient has a finite number of singular terms. That is, at a certain step, the renormalization of the mass becomes unnecessary. In our case, it was shown that $M_{r,3}^a\stackrel{\text{s.p.}}{=}0$, see formula (\ref{30-82}). This is due to the super-renormalizability of the theory. The linear coefficient $t_{r,i}^a$ has a similar property. For dimensional reasons, it can be argued that $t^a_{r,i}\stackrel{\text{s.p.}}{=}0$ for all $i\geqslant 5$. However, in the fourth loop, the singularities still appear and they are logarithmic in nature. This means that an ansatz for the fourth correction should be taken as
\begin{equation}\label{30-88}
t^a_{r,4}\stackrel{\text{s.p.}}{=}\sum_{k=1}^NL^kt^a_{r,4,k},
\end{equation}
where $N\in\mathbb{N}$ and $L=\ln(\Lambda/\sigma)$.

\begin{figure}[h]
	\center{\includegraphics[width=0.36\linewidth]{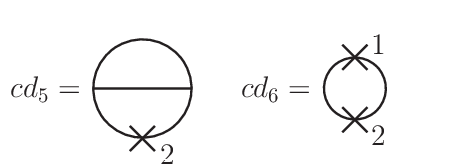}}
	\caption{The counter diagrams $cd_5$ and $cd_6$.}
	\label{30-pic-5}
\end{figure}

As mentioned earlier, singularities $L^{k+1}$ with $k>0$ can be calculated using a simpler method of differentiating the relations for the coefficients of the renormalization constants with respect to the auxiliary parameter $\sigma$. In the four-loop order we have
\begin{equation}\label{30-89}
-\sigma\frac{d}{d\sigma}\Bigg(\sum_{k=1}^NL^kt^a_{r,4,k}S_a[B]\Bigg)=-\sigma\frac{d}{d\sigma}(\text{4-loop diagrams and counter diagrams}).
\end{equation}
It is clear that the derivative with respect to $\sigma$ leads to zero all diagrams and counter diagrams that do not contain the two-loop coefficient $M^{ab}_{r,2}$. This means that only two non-trivial contributions remain. Therefore, relation (\ref{30-89}) is rewritten as
\begin{equation}\label{30-90}
\sum_{k=2}^NkL^{k-1}t^a_{r,4,k}=-\sigma\frac{d}{d\sigma}\bigg(-\frac{1}{4}cd_5+\frac{1}{2}cd_6\bigg)+\tilde{\kappa}_5,
\end{equation}
where $\tilde{\kappa}_5$ subtracts a singular density independent of the background field, and the counter  diagrams $cd_5$ and $cd_6$ are shown in Fig. \ref{30-pic-5}.

It is clear that the derivative on the right hand side of (\ref{30-90}) acts on $M^{ab}_{r,2}$, therefore, for convenience, we introduce the auxiliary notation $\mu^{ab}=-\sigma\partial_\sigma M^{ab}_{r,2}$. Considering the fact that $M^{ab}_{r,2}$ is linear in $L$, see (\ref{30-42}), then $\mu^{ab}$ does not depend on $\Lambda$ and is finite. 

Let us use notations (\ref{30-30}), (\ref{30-43}), and (\ref{30-45}). Then the right hand side of the equality can be rewritten as
\begin{equation*}\label{30-91}
-\frac{1}{4}\int_{\mathbb{R}^5\times\mathbb{R}^5}\mathrm{d}^5x\mathrm{d}^5y\,\Big(G^\Lambda_{ab}(x,y)\rho^{ac}_1(x,y)\mu^{bc}-G^\Lambda_{ab}(x,y)\rho^{ac}_2(x,y)\mu^{bc}\Big)+\tilde{\kappa}_5.
\end{equation*}
We immediately note that in the first term, only the following parts of the function $\rho^{ac}_1(x,y)$ can lead to singularities
\begin{equation*}\label{30-92}
\int_{\mathbb{R}^5}\mathrm{d}^5z\,g^{aa_1a_2}\Big(R^\Lambda_0(x-z)\Big)^2g^{a_1a_2b_3}G^\Lambda_{b_3b}(z,y)
\end{equation*}
and
\begin{equation*}\label{30-921}
-\int_{\mathbb{R}^5}\mathrm{d}^5z\,g^{aa_1a_2}R^\Lambda_0(x-z)R^\Lambda_1(x-z)g^{a_2b_1a_3}B_{b_1}(x)g^{a_1a_3b}R^\Lambda_0(z-y).
\end{equation*}
Let us use transformation (\ref{30-49}). Then the part with $\mathrm{A}(0)$ cancels the analogous (but with opposite sign) singularity from $\rho^{ac}_2(x,y)$, and the answer includes parts containing $\hat{R}^\Lambda_0(\cdot)$ and $\hat{R}^\Lambda_1(\cdot)$. In terms of the functions (\ref{30-51}), (\ref{30-53}), and (\ref{30-55}), the required contributions are written out as
\begin{multline*}\label{30-93}
-\frac{1}{4}\int_{\mathbb{R}^5\times\mathrm{B}_{1/\sigma}}\mathrm{d}^5x\mathrm{d}^5y\,\Big(R^\Lambda_0(y)\rho^{ab}_5(x,y+x)\mu^{ab}-R^\Lambda_{1}(y)g^{aa_1c}B_{a_1}(x)\rho^{ab}_3(x,y+x)\mu^{cb}+\\+R^\Lambda_0(y)\rho^{ab}_7(x,y+x)\mu^{ab}
\Big),
\end{multline*} 
and using the formulas (\ref{30-61}), (\ref{30-64}), and (\ref{30-65}), they are transformed as follows
\begin{equation*}\label{30-94}
-\frac{1}{4}S_{a_1}[B]\Big(-\mathrm{I}_8(\Lambda,\sigma)h_1^{ab}g^{ba_1c}\mu^{ca}-\mathrm{I}_5(\Lambda,\sigma)g^{ba_1c}h^{ab}_1\mu^{ca}-2\mathrm{I}_9(\Lambda,\sigma)g_1^{ba_1c}\mu^{cb}\Big).
\end{equation*}

Using the asymptotic expansion for auxiliary integrals (\ref{30-p24}), we finally obtain that in formula (\ref{30-88}) the parameter $N=2$ and the fourth coefficient has the form
\begin{equation}\label{30-95}
t^a_{r,4}=-\frac{L^2\big(2h^{a_1b}_1g^{bac}h_3^{ca_1}-
	h^{a_1b}_1g^{bac}h_2^{ca_1}
	-4g_1^{ba_1c}h_3^{cb}+
	2g_1^{ba_1c}h_2^{cb}
	\big)}{2^53^2(16\pi^2)^4}+Lt^a_{r,4,1}+\tilde{t}^a_{r,4},
\end{equation}
where $\tilde{t}^a_{r,4}$ is a constant. Thus, the quadratic logarithmic contribution to the fourth coefficient was found. The term proportional to the first power of the logarithm $L$ must be found by direct calculations of the fourth correction in full.

\section{Conclusion}

In this paper, the three-loop correction to the quantum action for the scalar cubic model in the case of a general formulation has been studied. The main results are formulated in Theorem \ref{30-th-1} in Section \ref{30:sec:prop}. They include answers for the coefficients of the renormalization constants in the first three loops, a partial answer for the four-loop term, the discussion of the absence of non-local terms, as well as a consequence of the super-renormalizability of the theory, which consists in a finite number of singular coefficients. After the theorem, some interesting special cases are presented.

It is worth noting separately that in order to renormalize the mass parameter, it was actually enough to consider the first two loops, since there are no singular contributions starting from the third loop. This fact is actually easy to verify using the analysis of dimension. The parameter for the linear term acquires singular contributions up to and including the fourth correction. In particular, using differentiation by the auxiliary parameter, the correction from the fourth loop proportional to $L^2$ was investigated, see Section \ref{30:sec:next-cor} and formula \eqref{30-95}.

\vspace{2mm}
\textbf{Acknowledgements.} The work is supported by the Ministry of Science and Higher Education of the Russian
Federation, grant 075-15-2022-289. Also, A.V.Ivanov is supported by the Foundation for the Advancement of Theoretical Physics and Mathematics "BASIS", grant "Young Russian Mathematics".

\vspace{2mm}
The authors express special gratefulness to K.A.Ivanov for creating comfortable and stimulating conditions for writing the work.

\vspace{2mm}
\textbf{Data availability statement.} Data sharing not applicable to this article as no datasets were generated or analysed during the current study.

\vspace{2mm}
\textbf{Conflict of interest statement.} The authors state that there is no conflict of interest.

\section{Appendix}
\label{30:sec:appen}

Let us consider the calculation of singular parts of auxiliary integrals. To do this, recall the relation connecting the Green's function for the free Laplace operator with its homogenized version
\begin{equation}\label{30-p1}
\frac{3}{8\pi^2}
\int_{\mathrm{S}^4}\mathrm{d}^4\hat{x}\,R_0(r\hat{x}+y)=\frac{1}{8\pi^2}\begin{cases}
\,\,r^{-3}, &|y|\leqslant r;\\
|y|^{-3}, &|y|>r,
\end{cases}
\end{equation}
where we have used the integration over the four-dimensional unit sphere $\mathrm{S}^4$ with the center at the origin and with the standard measure. Here $\hat{x}=x/|x|\in\mathrm{S}^4\subset \mathbb{R}^5$ and $r>0$.

Using relation (\ref{30-p1}), the following series of equalities can be verified by direct integration
\begin{equation}\label{30-p2}
\tilde{c}_2=\int_{\mathrm{B}_{1/\sigma}\times \mathrm{B}_{1/\sigma}}\mathrm{d}^5x\mathrm{d}^5y\,R_0(x)R_0(x+y)R_0(y)=\frac{1}{72\pi^2\sigma},
\end{equation}
\begin{equation}\label{30-p3}
R^\Lambda_1(y)\Big|_{\Lambda \to+\infty}=\int_{\mathrm{B}_{1/\sigma}}\mathrm{d}^5x\,R_0(x)R_0(x+y)=\frac{1}{24\pi^2}\begin{cases}
\frac{3}{2|y|}-\sigma, &|y|\leqslant 1/\sigma;\\
\,\,\frac{1}{2|y|^3\sigma^2}, &|y|>1/\sigma,
\end{cases}
\end{equation}
\begin{align}\nonumber
R^\Lambda_2(z)\Big|_{\Lambda\to+\infty}=&~2\int_{\mathrm{B}_{1/\sigma}\times \mathrm{B}_{1/\sigma}}\mathrm{d}^5x\mathrm{d}^5y\,R_0(x)R_0(x+y)R_0(y+z)-2\tilde{c}_2\\
\label{30-p4}=&~\frac{1}{36\pi^2}
\begin{cases}
-\frac{9|z|}{8}+\frac{3}{10}\sigma|z|^2, &|z|\leqslant 1/\sigma;\\
\,\,\,\,\frac{7}{40|z|^3\sigma^4}-\frac{1}{\sigma}, &|z|>1/\sigma.
\end{cases}
\end{align}
They are useful in calculating singular parts for auxiliary quantities, in particular for (\ref{30-36}) and (\ref{30-37}). Indeed, it can be noted that the functions $R^\Lambda_1(x)$ and $R^\Lambda_2(x)$ in the integrals $\mathrm{I}_2^{\phantom{1}}(\Lambda,\sigma)$ and $\mathrm{I}_3^{\phantom{1}}(\Lambda,\sigma)$ can be replaced with the limit values at $\Lambda\to+\infty$. This is due to the fact that the transitions 
\begin{equation*}\label{30-p5}
R^\Lambda_i(x)\to R^{+\infty}_i(x),\,\,\,\text{for}\,\,\,i=2,3,
\end{equation*}
can be performed by adding a function with a compact support to each $R^\Lambda_0$ in (\ref{30-23}) and (\ref{30-24}). Therefore, the integrals
\begin{equation*}\label{30-p6}
\int_{\mathrm{B}_{1/\sigma}}\mathrm{d}^5x\,R^\Lambda_0(x)\Big(\big(R^\Lambda_1(x)\big)^2-\big(R^{+\infty}_1(x)\big)^2\Big)\,\,\,\,\,\,\text{and} \,\,\,\,\,\,\int_{\mathrm{B}_{1/\sigma}}\mathrm{d}^5x\,\big(R^\Lambda_0(x)\big)^2\big(R^\Lambda_2(x)-R^{+\infty}_2(x)\big),
\end{equation*}
as functions of the parameter $\Lambda$, are finite in the transition $\Lambda\to+\infty$. Thus, using equalities (\ref{30-p3}) and (\ref{30-p4}), we obtain
\begin{equation}\label{30-p7}
\mathrm{I}_2(\Lambda,\sigma)=\int_{1/\Lambda}^{1/\sigma}\mathrm{d}r\,r^4\frac{8\pi^2}{3}\frac{1}{8\pi^2r^3}\frac{1}{2^8\pi^4r^2}+\mathcal{O}(1)=\frac{L}{3(16\pi^2)^2}+\mathcal{O}(1),
\end{equation}
\begin{equation}\label{30-p8}
\mathrm{I}_3(\Lambda,\sigma)=\int_{1/\Lambda}^{1/\sigma}\mathrm{d}r\,r^4\frac{8\pi^2}{3}\frac{1}{2^6\pi^4r^6}\bigg(-\frac{r}{32\pi^2}\bigg)+\mathcal{O}(1)=-\frac{L}{3(16\pi^2)^2}+\mathcal{O}(1).
\end{equation}
Let us move on to the consideration of the integral $\mathrm{I}_1(\Lambda,\sigma)$. It contains a power-law singularity, so the analysis procedure is different. Let us firstly write out the explicit formula
\begin{equation*}\label{30-p9}
\mathrm{I}_1(\Lambda,\sigma)=\int_{\mathrm{B}_{1/\sigma}\times\mathrm{B}_{1/\sigma}}\mathrm{d}^5x\mathrm{d}^5y\,\big(R^\Lambda_0(x)\big)^2R^\Lambda_0(y)R^\Lambda_0(y+x).
\end{equation*}
Note that in the integral over the variable $x$, the integration domain $\mathrm{B}_{1/\sigma}$ can be replaced by $\mathbb{R}^5$, since in this case we add a finite integral over $\mathbb{R}^5\setminus\mathrm{B}_{1/\sigma}$, which does not contain singularities. Additionally, we add and subtract a similar domain for the second integration, then
\begin{equation}\label{30-p10}
\mathrm{I}_1(\Lambda,\sigma)=\int_{\mathbb{R}^5}\mathrm{d}^5x\,\big(R^\Lambda_0(x)\big)^2\Bigg(\int_{\mathbb{R}^5}\mathrm{d}^5y\,R^\Lambda_0(y)R^\Lambda_0(y+x)-\int_{\mathbb{R}^5\setminus \mathrm{B}_{1/\sigma}}\mathrm{d}^5y\,R^\Lambda_0(y)R^\Lambda_0(y+x)\Bigg)+\mathcal{O}(1).
\end{equation}
Next, we note that the integral
\begin{equation}\label{30-p11}
\int_{\mathbb{R}^5}\mathrm{d}^5x\,\big(R^\Lambda_0(x)\big)^2\Bigg(\int_{\mathbb{R}^5\setminus \mathrm{B}_{1/\sigma}}\mathrm{d}^5y\,\Big[\big(R^\Lambda_0(y)\big)^2-R^\Lambda_0(y) R^\Lambda_0(y+x)\Big]\Bigg)
\end{equation}
converges and is finite when removing the regularization $\Lambda\to+\infty$. Indeed, to prove it, it is enough to note that in the region $|y|>1/\sigma$ and $|x|\leqslant1/\sigma_1$ for $\Lambda>\sigma$, where a fixed parameter $\sigma_1>0$ satisfies the inequality $1/\sigma_1+1/\Lambda\leqslant1/\sigma$, the relations hold
\begin{equation*}
R_0^{\Lambda}(y)=R_0^{\phantom{1}}(y),\,\,\,
\int_{\mathrm{S}^4}\mathrm{d}^4\hat{x}\,R_0^\Lambda(y+|x|\hat{x})=R_0^{\phantom{1}}(y),
\end{equation*}
and, thus, the following equality holds
\begin{equation*}
\int_{\mathrm{B}_{1/\sigma_1}}\mathrm{d}^5x\,\big(R^\Lambda_0(x)\big)^2\Bigg(\int_{\mathbb{R}^5\setminus \mathrm{B}_{1/\sigma}}\mathrm{d}^5y\,\Big[\big(R^\Lambda_0(y)\big)^2-R^\Lambda_0(y) R^\Lambda_0(y+x)\Big]\Bigg)=0.
\end{equation*}
This means that the integral \eqref{30-p11} tends to a finite number 
\begin{equation*}
	\int_{\mathbb{R}^5\setminus \mathrm{B}_{1/\sigma}}\mathrm{d}^5x\,\big(R_0(x)\big)^2\Bigg(\int_{\mathbb{R}^5\setminus \mathrm{B}_{1/\sigma}}\mathrm{d}^5y\,\Big[\big(R_0(y)\big)^2-R_0(y) R_0(y+x)\Big]\Bigg),
\end{equation*}
when removing the regularization
Hence, introducing the auxiliary definition
\begin{equation*}\label{30-p12}
\alpha_1(\mathbf{f})=\int_{\mathbb{R}^5}\mathrm{d}^5x\,\big(R^1_0(x)\big)^2\int_{\mathbb{R}^5}\mathrm{d}^5y\,R^1_0(y)R^1_0(y+x)
\end{equation*}
and scaling the variables $x\to x/\Lambda$ and $y\to y/\Lambda$ in the first term of (\ref{30-p10}), we get
\begin{align}
\mathrm{I}_1(\Lambda,\sigma)&=\Lambda^2\alpha_1(\mathbf{f})-\int_{\mathbb{R}^5}\mathrm{d}^5x\,\big(R^\Lambda_0(x)\big)^2\int_{\mathbb{R}^5\setminus \mathrm{B}_{1/\sigma}}\mathrm{d}^5y\,\big(R^\Lambda_0(y)\big)^2+\mathcal{O}(1)=\nonumber
\\
&=\Lambda^2\alpha_1(\mathbf{f})-\mathrm{A}(0)\big(\mathrm{A}(0)-\mathrm{A}(\sigma)\big)+\mathcal{O}(1).\label{30-p14}
\end{align}

Let us study singular terms of auxiliary integrals (\ref{30-60})--(\ref{30-65}) for the third correction. Let us start with $\mathrm{I}_4(\Lambda,\sigma)$. In this case, the parameter $\sigma$ appears in two places: during the final integration, see (\ref{30-60}), and in the definition of the function $R^\Lambda_1(\cdot)$, see (\ref{30-23}). In both cases, $\sigma$ is responsible for the radius $1/\sigma$ of the ball $\mathrm{B}_{1/\sigma}$, over which the integration is performed. Let us prove that it is enough to consider the case of $\sigma\to+0$.

First, we show that it is possible to replace $\mathrm{B}_{1/\sigma}$ with $\mathbb{R}^5$ in the definition of $R^\Lambda_1(\cdot)$. Consider the difference of the form
\begin{equation*}\label{30-p15}
\mathrm{I}_4(\Lambda,\sigma)-\int_{\mathrm{B}_{1/\sigma}}\mathrm{d}^5x\, \hat{R}^\Lambda_0(x)\Big(\hat{R}^\Lambda_1(x)\big|_{\sigma\to+0}\Big)=\int_{\mathrm{B}_{1/\sigma}}\mathrm{d}^5x\,\hat{R}^\Lambda_0(x)\Big(\hat{R}^\Lambda_1(x)-\hat{R}^\Lambda_1(x)\big|_{\sigma\to+0}\Big)\stackrel{\text{s.p.}}{=}0,
\end{equation*} 
where the facts that the combination $R^\Lambda_1(x)-R^\Lambda_1(x)|_{\sigma\to+0}$ does not contain singularities and is continuous were used. In particular, it has a finite limit at $x=0$.

Let us now consider the second parameter $\sigma$. We start with the formula for $\hat{R}^\Lambda_0(\cdot)$. Taking into account definition (\ref{30-49}) and formula (\ref{30-p1}), for $|x|>1/\Lambda$ the function can be written as
\begin{equation}\label{30-p16}
\hat{R}^\Lambda_0(x)=\int_{\mathbb{R}^5}\mathrm{d}^5z\,\big(R^\Lambda_0(z)\big)^2\big(R^\Lambda_0(z+x)-R_0(z+x)\big)-\mathrm{A}(0)\big(R^\Lambda_0(x)-R_0(x)\big)-\frac{1}{(16\pi^2)^2|x|^4},
\end{equation}
where for the region $|x|>1/\Lambda$ we have used the chain of equalities
\begin{multline*}
	\hat{R}_0^{\phantom{1}}(x)=\int_{\mathbb{R}^5}\mathrm{d}^5z\,\big(R^\Lambda_0(z)\big)^2\big(R_0^{\phantom{1}}(z+x)-R_0^{\phantom{1}}(x)\big)=\int_{\mathbb{R}^5}\mathrm{d}^5z\,\big(R^\Lambda_0(z)\big)^2\Big(R_0^{1/|z|}(x)-R_0^{\phantom{1}}(x)\Big)=
	\\=
	\frac{1}{(8\pi^2)^3}\int_{\mathbb{R}^5\setminus \mathrm{B}_{|x|}}\mathrm{d}^5z\,\bigg(\frac{1}{|z|^9}-\frac{1}{|x|^3|z|^6}\bigg)=
	-\frac{1}{(16\pi^2)^2|x|^4}.
\end{multline*}
For $|x|\leqslant 1/\Lambda$, the following relation is valid
\begin{equation*}
	\hat{R}_0^{\phantom{1}}(x)=
	\frac{\Lambda^4}{3(8\pi^2)^2}\int_{\Lambda|x|}^1\mathrm{d}t\,\bigg(t-\frac{t^4}{\Lambda^3|x|^3}\bigg)
	\bigg(\mathbf{f}\big(t^2\big)+1\bigg)^2-\frac{\Lambda^4}{(16\pi^2)^2}.
\end{equation*}
Note that $\mathrm{supp}\big(R^\Lambda_0(\cdot)-R_0(\cdot)\big)\subset\mathrm{B}_{1/\Lambda}$, so $\hat{R}^\Lambda_0(x)$ for large values of the variable $|x|$ has the behavior $|x|^{-4}$ in the main order. Similarly, it is checked that $\hat{R}^\Lambda_1(x)$ has the behavior $|x|^{-2}$. Therefore, replacing $\mathrm{B}_{1/\sigma}$ with $\mathbb{R}^5$ in (\ref{30-60}), only a finite part is added. Thus, it was shown that
\begin{multline}\label{30-p17}
\mathrm{I}_4(\Lambda,\sigma)=\Lambda\int_{\mathbb{R}^5\times\mathbb{R}^5}\mathrm{d}^5x\mathrm{d}^5z\,\bigg(\int_{\mathbb{R}^5}\mathrm{d}^5y\,\big(R^1_0(y)\big)^2R^1_0(y+x)-\mathrm{A}(0)R^1_0(x)\bigg)R^1_0(x+z)\times\\
\times\bigg(\int_{\mathbb{R}^5}\mathrm{d}^5u\,\big(R^1_0(u)\big)^2R^1_0(u+z)-\mathrm{A}(0)R^1_0(z)\bigg)=\Lambda\alpha_2(\mathbf{f}),
\end{multline}
where the variables were scaled and the following relation was used
\begin{equation}\label{30-p18}
\hat{R}^\Lambda_1(x)\Big|_{\sigma\to+0}=\int_{\mathbb{R}^5}\mathrm{d}^5y\,\hat{R}^\Lambda_0(y)R^\Lambda_0(x+y).
\end{equation}

Let us move on to the integral $\mathrm{I}_5(\Lambda,\sigma)$ from (\ref{30-61}). Using the previous arguments of replacing $\mathrm{B}_{1/\sigma}$ with $\mathbb{R}^5$ in the definition of $R^\Lambda_1(\cdot)$, the function under study is rewritten in the form
\begin{equation}\label{30-p19}
\mathrm{I}_5(\Lambda,\sigma)=\int_{\mathrm{B}_{\Lambda/\sigma}\times\mathbb{R}^5}\mathrm{d}^5x\mathrm{d}^5y\,\hat{R}^1_0(x)R^1_0(x+y)R^1_0(y)+\mathcal{O}(1),
\end{equation}
where additionally in the first term the scaling of $x\to x/\Lambda$ and $y\to y/\Lambda$ was done. Next, note that the functions $R^1_0(x+y)$ and $R^1_0(y)$ can be replaced with $R_0(x+y)$ and $R_0(y)$, since the replacement is performed by adding a function with a compact support that changes the integral (\ref{30-p19}) by a finite (non-singular) value. Therefore, using formulas (\ref{30-p3}) and (\ref{30-p16}) and the fact that the singularity in (\ref{30-p19}) is logarithmic, we have
\begin{align}\nonumber
\mathrm{I}_5(\Lambda,\sigma)\stackrel{\text{s.p.}}{=}&\,L\Bigg(\Lambda\frac{d}{d\Lambda}\int_{\mathrm{B}_{\Lambda/\sigma}\times\mathbb{R}^5}\mathrm{d}^5x\mathrm{d}^5y\,\hat{R}^1_0(x)R_0^{\phantom{1}}(x+y)R_0^{\phantom{1}}(y)\Bigg)\Bigg|_{\Lambda\to+\infty}=\\
=&~L\Bigg(\frac{8\pi^2}{3}\bigg(\frac{\Lambda}{\sigma}\bigg)^5\hat{R}^1_0(x)R^{+\infty}_1(x)\Big|_{\sigma\to+0,\,x=\Lambda/\sigma}\Bigg)\Bigg|_{\Lambda\to+\infty}=-\frac{L}{6(16\pi^2)^2}.\label{30-p21}
\end{align}

It is clear that the integral $\mathrm{I}_7(\Lambda,\sigma)$ is equal to the first term from (\ref{30-p19}) and, therefore, has the answer (\ref{30-p21}). Also, taking into account formula (\ref{30-p18}), it is not difficult to show that the integral $\mathrm{I}_8(\Lambda,\sigma)$ differs from $\mathrm{I}_7(\Lambda,\sigma)$ by a finite value. Indeed, the following chain of relations is valid
\begin{align}
\mathrm{I}_8(\Lambda,\sigma)=&\int_{\mathrm{B}_{1/\sigma}\times\mathbb{R}^5}\mathrm{d}^5x\mathrm{d}^5y\,R^\Lambda_0(x)R^\Lambda_0(x+y)\hat{R}^\Lambda_0(y)+\mathcal{O}(1)=\label{30-p22}\\
=&~\mathrm{I}_7(\Lambda,\sigma)+\int_{\mathrm{B}_{1/\sigma}\times(\mathbb{R}^5\setminus \mathrm{B}_{1/\sigma})}\mathrm{d}^5x\mathrm{d}^5y\,R^\Lambda_0(x)R^\Lambda_0(x+y)\hat{R}^\Lambda_0(y)-\\
-&\int_{(\mathbb{R}^5\setminus \mathrm{B}_{1/\sigma})\times \mathrm{B}_{1/\sigma}}\mathrm{d}^5x\mathrm{d}^5y\,R^\Lambda_0(x)R^\Lambda_0(x+y)\hat{R}^\Lambda_0(y)+\mathcal{O}(1)=\mathrm{I}_7(\Lambda,\sigma)+\mathcal{O}(1).\label{30-p23}
\end{align}
Then the general answer can be written as
\begin{equation}\label{30-p24}
\mathrm{I}_5(\Lambda,\sigma)\stackrel{\text{s.p.}}{=}\mathrm{I}_7(\Lambda,\sigma)\stackrel{\text{s.p.}}{=}\mathrm{I}_8(\Lambda,\sigma)\stackrel{\text{s.p.}}{=}-\frac{L}{6(16\pi^2)^2}.
\end{equation}

Let us move on to the integral $\mathrm{I}_6(\Lambda,\sigma)$. In this case, we use the general logic proposed in the analysis of $\mathrm{I}_4(\Lambda,\sigma)$, that is, we reduce the domains of integration of the form $\mathrm{B}_{1/\sigma}$ to $\mathbb{R}^5$. Let us start with the parameter in $R^\Lambda_1(x)$. Unlike the integral $\mathrm{I}_4(\Lambda,\sigma)$, in which $R^\Lambda_1(x)$ enters through $\hat{R}^\Lambda_1(x)$ and, therefore, is constructed through the difference according to (\ref{30-49}), in the integral $\mathrm{I}_6(\Lambda,\sigma)$ we can not replace $\mathrm{B}_{1/\sigma}$ with $\mathbb{R}^5$ without preserving the additional difference (at least its main part). In this case, the first correction term must be retained. In other words, we can make a replacement in the form
\begin{equation}\label{30-p25}
R^\Lambda_1(x-z)\to\int_{\mathbb{R}^5}\mathrm{d}^5y\,R^\Lambda_0(x-y)R^\Lambda_0(y-z)+\big(\mathrm{A}(\sigma)-\mathrm{A}(0)\big),
\end{equation}
and obtain
\begin{equation}\label{30-p26}
\mathrm{I}_6(\Lambda,\sigma)=\int_{\mathrm{B}_{1/\sigma}\times\mathbb{R}^5}\mathrm{d}^5x\mathrm{d}^5z\,\hat{R}^\Lambda_0(x)R^\Lambda_0(x-z)\Big(R^\Lambda_1(x-z)\big|_{\sigma\to+0}\Big)R^\Lambda_0(z)+\big(\mathrm{A}(\sigma)-\mathrm{A}(0)\big)\mathrm{I}_7(\Lambda,\sigma)+\mathcal{O}(1).
\end{equation}
Further, given the decreasing property of the function used above, we can replace $\mathrm{B}_{1/\sigma}$ with $\mathbb{R}^5$ in the first term of (\ref{30-p26}). Then, by scaling and using answer (\ref{30-p24}), we get
\begin{equation}\label{30-p27}
\mathrm{I}_6(\Lambda,\sigma)=\Lambda\alpha_3(\mathbf{f})-\frac{L\big(\mathrm{A}(\sigma)-\mathrm{A}(0)\big)}{6(16\pi^2)^2}+\mathcal{O}(1),
\end{equation}
where we have introduced the notation
\begin{equation*}\label{30-p28}
\alpha_3(\mathbf{f})=\int_{\mathbb{R}^5\times\mathbb{R}^5\times\mathbb{R}^5}\mathrm{d}^5x\mathrm{d}^5y\mathrm{d}^5z\,\hat{R}^1_0(x)R^1_0(x-z)R^1_0(x-y)R^1_0(y-z)R^1_0(z).
\end{equation*}

Moving on to the last function from (\ref{30-65}), we immediately note that in $R^\Lambda_1(x-z)$, we can select the point $\sigma=+0$. Despite the fact that $R^\Lambda_1(x-z)$ is not included into a difference, as it was in $\mathrm{I}_4(\Lambda,\sigma)$, we do not need to keep in mind the correction term (as in (\ref{30-p25})), because $\mathrm{I}_9(\Lambda,\sigma)$ contains a logarithmic singularity. Unlike the situation in $\mathrm{I}_6(\Lambda,\sigma)$, which contains a power-law singularity, and correction plays an essential role in it. Thus, by making an additional permutation of the multipliers, we have
\begin{equation*}\label{30-p29}
\mathrm{I}_9(\Lambda,\sigma)=\int_{\mathrm{B}_{1/\sigma}\times\mathbb{R}^5}\mathrm{d}^5x\mathrm{d}^5z\,R^\Lambda_0(x)R^\Lambda_0(x+z)R^\Lambda_0(z)\Big(R^\Lambda_1(z)\big|_{\sigma\to+0}\Big)+\mathcal{O}(1).
\end{equation*}
Next, using a procedure similar to (\ref{30-p22})--(\ref{30-p23}), we reduce the integration domain related to the variable $z$ from $\mathbb{R}^5$ to $\mathrm{B}_{1/\sigma}$, and the integral with $x$ is extended from $\mathrm{B}_{1/\sigma}$ to $\mathbb{R}^5$. As a result, we get
\begin{equation}\label{30-p30}
\mathrm{I}_9(\Lambda,\sigma)=\int_{\mathrm{B}_{1/\sigma}\times\mathbb{R}^5}\mathrm{d}^5z\mathrm{d}^5x\,R^\Lambda_0(z)\Big(R^\Lambda_1(z)\big|_{\sigma\to+0}\Big)R^\Lambda_0(x+z)R^\Lambda_0(x)+\mathcal{O}(1).
\end{equation}
Note that in the integral with the variable $x$, the function $R^\Lambda_1(z)$ was obtained at $\sigma =0$. By changing the first term in (\ref{30-p30}) to a finite value, we move from $R^\Lambda_1(z)$ to $R^{+\infty}_1(z)$. Finally,  after scaling the variables, we have
\begin{equation}\label{30-p31}
\mathrm{I}_9(\Lambda,\sigma)\stackrel{\text{s.p.}}{=}L\Bigg(\Lambda\frac{d}{d\Lambda}\int_{\mathrm{B}_{1/\sigma}}\mathrm{d}^5z\,R^1_0(z)\Big(R^{+\infty}_1(z)\big|_{\sigma\to+0}\Big)^2\Bigg)\Bigg|_{\Lambda=+\infty}=\frac{L}{3(16\pi^2)^2}.
\end{equation}
In addition, we note that $\mathrm{I}_9(\Lambda,\sigma)\stackrel{\text{s.p.}}{=}\mathrm{I}_2(\Lambda,\sigma)$.

Let us proceed to the calculation of integrals arising in the diagram $d_3$, that is, $\mathrm{I}_{10}(\Lambda,\sigma)$ and $\mathrm{I}_4(\Lambda,\sigma)$ from (\ref{30-76}) and (\ref{30-77}). We start with the latter, since it will be used in $\mathrm{I}_{10}(\Lambda,\sigma)$. Equation (\ref{30-77}) contains the explicit formula, but it is not convenient for calculating. Let us make some transformations. First, we reduce the integration domain for the variable $y$ from $\mathbb{R}^5$ to $\mathrm{B}_{1/\sigma}$, and extend the domain related to the variable $z$ to $\mathbb{R}^5$. At the same time, $\mathrm{I}_{11}(\Lambda,\sigma)$ is changed only by a finite value because the integrals
\begin{equation*}\label{30-p32}
\int_{\mathrm{B}_{1/\sigma}\times\mathbb{R}^5\times(\mathbb{R}^5\setminus \mathrm{B}_{1/\sigma})}
\mathrm{d}^5z\mathrm{d}^5x\mathrm{d}^5y\,
R^\Lambda_0(z-y)R^\Lambda_0(z-x)R^\Lambda_0(x-y)R^\Lambda_0(x)R^\Lambda_0(y),
\end{equation*}
\begin{equation*}\label{30-p33}
\int_{(\mathbb{R}^5\setminus \mathrm{B}_{1/\sigma})\times\mathbb{R}^5\times \mathrm{B}_{1/\sigma}}
\mathrm{d}^5z\mathrm{d}^5x\mathrm{d}^5y\,
R^\Lambda_0(z-y)R^\Lambda_0(z-x)R^\Lambda_0(x-y)R^\Lambda_0(x)R^\Lambda_0(y)
\end{equation*}
exist and are finite, including after the removal of the regularization. Next, we replace the four functions  $R^\Lambda_0(\cdot)$, except the one with the argument $y$, with the limit one $R_0(\cdot)$. In this case, the initial integral changes to a finite value, since the transformation was performed by adding functions with a compact support. Also, note that such transformations are acceptable, since the singularity is the weakest, that is, logarithmic. It is also easy to verify (make estimates) that the additional terms  from differences are finite. As a result, we get an answer in the form
\begin{equation}\label{30-p34}
\mathrm{I}_{11}(\Lambda,\sigma)=\int_{\mathbb{R}^5\times\mathbb{R}^5\times \mathrm{B}_{1/\sigma}}\mathrm{d}^5z\mathrm{d}^5x\mathrm{d}^5y\,R_0^{\phantom{1}}(z-y)R_0^{\phantom{1}}(z-x)R_0^{\phantom{1}}(x-y)R_0^{\phantom{1}}(x)R^\Lambda_0(y)+\mathcal{O}(1).
\end{equation}
Next, using formula (\ref{30-p3}) in the form
\begin{equation*}\label{30-p35}
\int_{\mathbb{R}^5}\mathrm{d}^5z\,R_0(z-y)R_0(z-x)=\frac{1}{16\pi^2|x-y|},
\end{equation*}
making the replacement $x\to x+y$ in (\ref{30-p34}), and applying relation (\ref{30-p1}), we can rewrite the integral as follows
\begin{align}
\mathrm{I}_{11}(\Lambda,\sigma)&=\int_{\mathbb{R}^5\times \mathrm{B}_{1/\sigma}}\mathrm{d}^5x\mathrm{d}^5y\,\frac{2}{(16\pi^2)^2|x|^4}\Bigg(\frac{1}{8\pi^2}\begin{cases}
|x|^{-3}, &|y|\leqslant |x|;\\
|y|^{-3}, & |y|>|x|.
\end{cases}\Bigg)R^\Lambda_0(y)+\mathcal{O}(1)=\nonumber
\\
&=\frac{L}{3(16\pi^2)^2}+\mathcal{O}(1).\label{30-p37}
\end{align}

Moving on to the integral $\mathrm{I}_{10}(\Lambda,\sigma)$ from (\ref{30-76}), we immediately note that it is convenient to consider the transition to $\sigma=0$. As in the integral $\mathrm{I}_6(\Lambda,\sigma)$, the parameter $\sigma$ in it is in the integration domain (external, see (\ref{30-76})), and in the function $R^\Lambda_1(\cdot)$. In the latter case, as already noted, it is necessary to save the correction term, that is, use the transition from (\ref{30-p25}). The domain of integration (\ref{30-76}) can simply be replaced with $\mathbb{R}^5$, since at infinity the integral is convergent and, therefore, the singular part does not change with such replacement. Thus, after additional scaling of the variables, we get
\begin{align}
\mathrm{I}_{10}(\Lambda,\sigma)=&~\Lambda\alpha_4(\mathbf{f})+\big(\mathrm{A}(\sigma)-\mathrm{A}(0)\big)\mathrm{I}_{11}(\Lambda,\sigma)+\mathcal{O}(1)\nonumber
\\
=&~\Lambda\alpha_4(\mathbf{f})+\frac{L\big(\mathrm{A}(\sigma)-\mathrm{A}(0)\big)}{3(16\pi^2)^2}+\mathcal{O}(1),\label{30-p39}
\end{align}
where
\begin{equation*}\label{30-p43}
\alpha_4(\mathbf{f})=\int_{\mathbb{R}^{5\times4}}\mathrm{d}^5x_1\mathrm{d}^5x_2\mathrm{d}^5x_3\mathrm{d}^5x_4\,R^1_0(x_1)R^1_0(x_2)R^1_0(x_1-x_2)R^1_0(x_1-x_3)R^1_0(x_2-x_3)R^1_0(x_3-x_4)R^1_0(x_4).
\end{equation*}

The paper also uses the fact (see the text before formula (\ref{30-39})) that $R^\Lambda(0)$ behaves like $\Lambda^{-1}$ when $\Lambda\to+\infty$. Let us show it. In this case, we use definitions (\ref{30-24}) and (\ref{30-25}), then
\begin{equation*}\label{30-p40}
R^\Lambda_2(0)=2\int_{\mathrm{B}_{1/\sigma}\times \mathrm{B}_{1/\sigma}}\mathrm{d}^5x\mathrm{d}^5y\,\Big(R^\Lambda_0(x)R^\Lambda_0(x+y)R^\Lambda_0(y)-R_0^{\phantom{1}}(x)R_0^{\phantom{1}}(x+y)R_0^{\phantom{1}}(y)\Big).
\end{equation*}
 Let us scale $x\to x/\Lambda$ and $y\to y/\Lambda$, then formula (\ref{30-40}) can be rewritten as
\begin{equation}\label{30-p41}
R^\Lambda_2(0)=\frac{2}{\Lambda}\int_{\mathrm{B}_{\Lambda/\sigma}\times \mathrm{B}_{\Lambda/\sigma}}\mathrm{d}^5x\mathrm{d}^5y\,\Big(R^1_0(x)R^1_0(x+y)R^1_0(y)-R_0^{\phantom{1}}(x)R_0^{\phantom{1}}(x+y)R_0^{\phantom{1}}(y)\Big).
\end{equation}
Note that the integrand is rewritten as a finite sum, each term of which contains at least one function $R^1_0(\cdot)-R_0(\cdot)$, the support of which lies in $\mathrm{B}_1$. Therefore, the integral in (\ref{30-p41}) converges at $\Lambda\to +\infty$. Then the answer looks like
\begin{equation}\label{30-p42}
R^\Lambda_2(0)=\frac{2}{\Lambda}\int_{\mathbb{R}^5\times \mathbb{R}^5}\mathrm{d}^5x\mathrm{d}^5y\,\Big(R^1_0(x)R^1_0(x+y)R^1_0(y)-R_0^{\phantom{1}}(x)R_0^{\phantom{1}}(x+y)R_0^{\phantom{1}}(y)\Big)+o(1/\Lambda).
\end{equation}

\end{document}